\documentclass[11pt,a4paper]{article}
\usepackage[utf8]{inputenc}
\usepackage{color}
\usepackage{tikz}
\usepackage{eurosym}
\usetikzlibrary{positioning}
\usepackage{amsmath}
\usepackage{amssymb}
\usepackage{mathtools}
\usepackage[colorlinks = true,
            linkcolor = blue,
            urlcolor  = blue,
            citecolor = blue,
            anchorcolor = blue]{hyperref}

\providecommand{\keywords}[1]
{
  \small	
  \textbf{\textit{Keywords:}} #1
}

\usepackage[authoryear]{natbib}
\usepackage{cancel}
\usepackage{multirow}
\usepackage{amsthm}
\newtheorem{acknowledgement*}{Acknowledgement}
\newtheorem{result}{Result}
\usepackage{comment}
\usepackage{authblk}
\usepackage{fancyhdr}
\usepackage{geometry}

\usepackage[normalem,normalbf]{ulem}
\usepackage{tcolorbox}

\title{Efficiency in European Air Traffic Management - A Fundamental Analysis of Data, Models, and Methods}
\author[1]{Thomas Standfuss}
\author[2]{Georg Hirte}
\author[3]{Michael Schultz}
\author[1]{Hartmut Fricke}
\affil[1]{TU Dresden, Institute of Logistics and Aviation}
\affil[2]{TU Dresden, Institute of Transport and Economics}
\affil[3]{Bundeswehr University Munich, Institute of Flight Systems}

\date{January 2023}

\begin{document}
\maketitle
\begin{abstract}
We systematically study cornerstones that must be solved to define an air traffic control benchmarking system based on a Data Envelopment Analysis. Primarily, we examine the appropriate decision-making units, what to consider and what to avoid when choosing inputs and outputs in the case that several countries are included, and how we can identify and deal with outliers, like the Maastricht Service Provider. We argue that Air Navigation Service Providers would be a good choice of decision units within the European context. Based on that, we discuss candidates for DEA inputs and outputs and emphasize that monetary values should be excluded. We, further suggest to use super-efficiency DEA for eliminating outliers.  In this context, we compare different DEA approaches and find that standard DEA is performing well.
\end{abstract}


\pagestyle{fancy}
\fancyhead{}
\setlength{\headheight}{13.6pt}
\fancyhead[C]{Journal of Air Transport Management}
\renewcommand{\footrulewidth}{0.4pt}

\fancyfoot{}
\fancyfoot[R]{\thepage}
\fancyfoot[L]{Efficiency in European Air Traffic Management}

\section{Background}
The provision of Air Navigation Services (ANS) in Europe has gained increasing attention from the pre- to the post-pandemic, both from an academic side and from policy decision-makers. The evaluation and optimization of air traffic management, which suffered huge volatility in demand for service provision during that period,  can be approached from two economic perspectives: The microscopic and macroscopic levels. Considering the first canvas, optimization aims to route individual or multiple flights through the airspace according to well-defined performance criteria (ref: SES Performance Scheme). The objective function is usually multi-criteria, i.e., the flight trajectory generated follows objectives with volatile weighting ranging from classical operating costs to a set of different emissions such as noise, CO2, and non-CO2 footprints from, e.g., contrails. \citep{Rosenow2018,Rosenow2020,Fricke2021b}. For the second canvas focusing on the macroscopic view, we shift from individual flights to traffic flows, tackling the overall performance of a system or its units -- in our decision, the ANS providers (ANSP). This comparison is essential in the case of monopolistic structures since pricing cannot be used as an efficiency criterion.\

The field of ANS operations is complex and multi-layered. Air traffic control companies consist of several operational units and corresponding decision-makers. Despite ANS being managed in different units, we refer to ANSPs because they grant safety in air traffic operations and have decision power regarding important inputs, e.g., air traffic controllers, ATCO. In addition, ANSP must formally comply with set regulations and related pan-European assessments (SES performance scheme, \cite{EUROCONTROL2021b}). Thus, from an economic point of view, we face a multidimensional input-output problem.

Performance assessment aims to improve the efficiency of the ATM system and identify the contribution of single stakeholders to it, particularly the ANSP. Such assessment is not intended to act in a sense of a blame culture. However, it shall motivate incentive-driven to improve operational processes and strategic investment and work out the responsibility of each participant. Therefore, as a first step, we need to calculate the ANSPs' current performance, in other words, which achieves the highest (rank of) productivity or efficiency. Previous investigations, e.g., conducted by EUROCONTROL, focus on benchmarking ANSPs within Europe \citep{EUROCONTROL2019g,EUROCONTROL2019h} or on comparing Europe with the US \citep{EUROCONTROL2019c,EUROCONTROL2019j}. However, as further elaborated in section 3.1, these official reports need to improve regarding the benchmarking methodology, used models and metrics, information content, and root cause analysis. This is common sense in both groups: operational experts and academic researchers. The latter has partially addressed these weaknesses, such as the single-input-single-output schemes, by applying alternative methods such as Data Envelopment Analyses (DEA) or Stochastic Frontier Analysis (SFA). 

However, there has yet to be a fundamental discussion regarding economic modeling and applying these methods, considering the specific characteristics of ANS. We aim to close this gap and focus on the non-parametric approach of Data Envelopment Analysis. It allows the parallel use of multiple inputs and outputs and does not require assumptions about functional relationships or error terms, as is the case when applying SFA. To sum up, we will address the following research questions in this paper:
\begin{enumerate}
    \item How to model non-parametrically the economic value chain for a benchmark of the European ANSP?
    \item Which ANSP units should be considered in such benchmarking? 
    \item How to identify and deal with outliers?
    \item Which DEA method is appropriate to benchmark Air Navigation Services?
\end{enumerate}

The first decision we have to make is to define the objective of the benchmark. It is either to improve efficiency\footnote{Efficiency either means to provide services at minimum costs or to produce a fixed output with a minimal deployment of resources (or vice versa). We focus on the latter.} by considering choices made by the assessed decision units (e.g., the sectors) and the upper-level decision units (e.g., the ANSP) or to focus only on the decision unit itself. We follow the second approach: Therefore, the benchmark is related to the decision power of the ANSP. Consequently, we only consider input and output variables that are related to the decision power of the ANSP. However, we do not include inputs chosen in other operational levels of the ANS structure (ACC-, sector group-, or sector-level, \cite{Standfuss2021}). 

In order to present our recommendations regarding a benchmarking procedure, we structure the paper as follows: Section 2 deals with the basics of efficiency assessment and gives an overview of studies in the context of aviation. Section 3 examines the air navigation services' environmental factors and particularities. Section 4 presents the application, the results, and the discussion regarding plausibility and robustness of our findings. Section 5 summarizes the results and provides an outlook.\

\section{Performance Assessment in ATM}
\subsection{Literature Review}

In the late 1990s, EUROCONTROL began evaluating the performance of air navigation services. Since 2003,it  has published the respective data and produced various reports. The best-known reports are the ATM Cost Effectiveness Report \citep{EUROCONTROL2019g} and the Performance Review Report \citep{EUROCONTROL2019h}. EUROCONTROL uses a two-dimensional productivity measure (one output divided by one input or costs divided by output). The goal is to rank the European service providers and identify influencing factors. Despite the results being intuitively easy to understand, the method needs improvement. First, more than a two-dimensional measure is required to reflect the complexity and heterogeneity of European ANSPs. Second, several studies have demonstrated significant areas for improvement in data and metrics \citep{Standfuss2021b, Fricke2021b, FABEC2020a}. Since the results also have high political relevance, improving the benchmarking scheme is relevant and necessary.\

Subsequently, more and more academic studies addressed the performance benchmarking of ANSPs, successively eliminating the methodological weaknesses of the official reports and thus gaining further insights into the mechanisms of efficiency drivers and blockers. The first study on ANSP benchmarking, which considers multiple, is represented by \citet{NERA2006}. The authors estimated a Cobb-Douglas cost function applying SFA with fixed and random effects models. \citet{Button2013} publishes a paper on the potential economic benefits of Functional Airspace Blocks (FABs). Using a bootstrapped DEA, he tests a model consisting of two cost-based inputs and three outputs (flight hours, airport movements, and delays). The DEA values are aggregated per FAB. A regression analysis tests the influence of different factors on the efficiency values. Unfortunately, the procedure is only superficially explained and justified. The same criticisms apply to \citet{Button2014}, which agrees in methodology and modeling with \citet{Button2013}. Based on the DEA method, \citet{Cujic2015} investigate the efficiency of European ANSPs in the years 2009-2011 using a model with two cost-based inputs and three outputs (delays, Composite Flight Hours, and total revenue). However, the results are not robust regarding efficiency scores or rankings. \citet{Neiva2014} applies both the DEA and the SFA for performance benchmarking, covering FAB and ANSP levels. The modeling does not differ from \citet{Button2013} or \citet{Button2014}. \

\citet{Arnaldo2014} also used DEA to measure the efficiency of 35 European ANSPs. The data contains the years 2001 to 2011. The authors provide a comprehensive description of the data used and the modeling. They tested different approaches in orientation (input vs.\ output) and returns to scale (constant\footnote{CRS} vs.\ variable\footnote{VRS}). The results show that too many factors were considered simultaneously within a DEA model: Almost half of the units are efficient in the VRS-DEA. \citet{Bilotkach2015} calculate the cost efficiency and productivity of European ANSPs using a Malmquist DEA model. Data from the years 2002 to 2011 were available for the study. The authors use `controlled flight hours' and `aircraft movements at the airport' as outputs, as well as gate-to-gate ATM/CNS costs as inputs. Furthermore, they implement input prices for controllers, other personnel, capital, and other resources to determine allocative efficiency.  

\citet{Adler2017} address the connection between performance and ownership of the ANSP. The authors differentiate whether they are state organizations or partially privatized companies. They use the SFA to estimate the production and cost function of 37 ANSPs based on nine years of data. As a result, ANSPs with public-private ownership with stakeholder participation achieved statistically significantly higher productivity and cost efficiency than a state-owned enterprise or government agency.

More recently, an academic group supporting Performance Review Board (PRB) identifies tremendous inefficiencies in European ANS and provides recommendations for regulation in reference period 3 (RP3) \citep{PRB2018}. However, the results and subsequent statements are at least debatable since modeling and application of the methods do not appropriately reflect operational determinants of ANSPs, as emphasized by \citep{Standfuss2020}. The findings further show the necessity of questioning the used data and metrics provided by EUROCONTROL.

To sum up, all studies mentioned show one or more areas for improvement. First, the models often do not reflect the relevant outputs and inputs of the ANSPs (see also section 3). Second, including too many factors in the benchmarking model usually leads to many units being reported as efficient. Third, the use of costs is very questionable, as wage effects often play a role and are exogenous to the ANSPs. Fourth, the biggest issue is the lack of methodological discussion. For example, \citet{Button2013} states that bootstrapping is necessary but do not provide an explanation or justification. Further, most studies use Data Envelopment Analysis but do not provide a comparison across different methods. Our contribution intends to close this gap.

More recent publications address specific aspects of efficiency drivers based on airspace structure or its management. One of the most prominent recommendations to increase efficiency is the consolidation of airspace units. The idea of merging airspaces is not new since EUROCONTROL introduced functional airspace blocks some twenty years ago. However, progress is marginal and the effectiveness of the FAB concept in the current allocation is to be questioned \citep{Standfuss2019f}. 

The Cadenza project achieved more promising results, showing significant cost savings by merging the management of airspaces that are not adjacent \citep{Starita2021}. Other studies deals with business models \citep{Buyle2020}, regulatory frameworks for cost-efficient ANSPs with increased capacity \citep{Adler2022}, or consequences of a potential privatization \citep{Buyle2022}. Although these studies gave fundamental insights in efficiency of ANS provision, the question of a fundamental benchmarking concept still needs to be answered.

\subsection{ANSPs as Natural Candidates for Benchmarking}
The main task of an ANSP is to avoid mid-air collisions. Therefore, Air Traffic Controllers (ATCOs) separate the traffic vertically and horizontally. European ANSPs deal mostly with commercial traffic using Instrument Flight Rules (IFR), which represents a share of 90\% of all controlled movements. Furthermore, this type of traffic is the one that is charged to finance the ANS services.\ 

The historical approach of managing airspace in accordance with national boundaries has led to an airspace structure in Europe characterized by a high level of spatial fragmentation. Thus, airspace boundaries and partly sectors were not determined based on the dynamic traffic demand/flows but mainly according to national territories. This current structure of European ATM may lead to inefficiencies caused by additional coordination efforts and inconsistencies between ANSPs’ strategies and capacity restrictions \citet{EUROCONTROL2019c, Standfuss2019f}. However, one should also note that this fragmentation of airspace also affects operations by ANSPs. As an example, procedures are influenced by the geographic environment as well as by the airspace structure. These unique determinants have to be considered by air traffic controllers.\ 

\begin{figure}[htb!]
    \centering
    \includegraphics[width=0.75\textwidth]{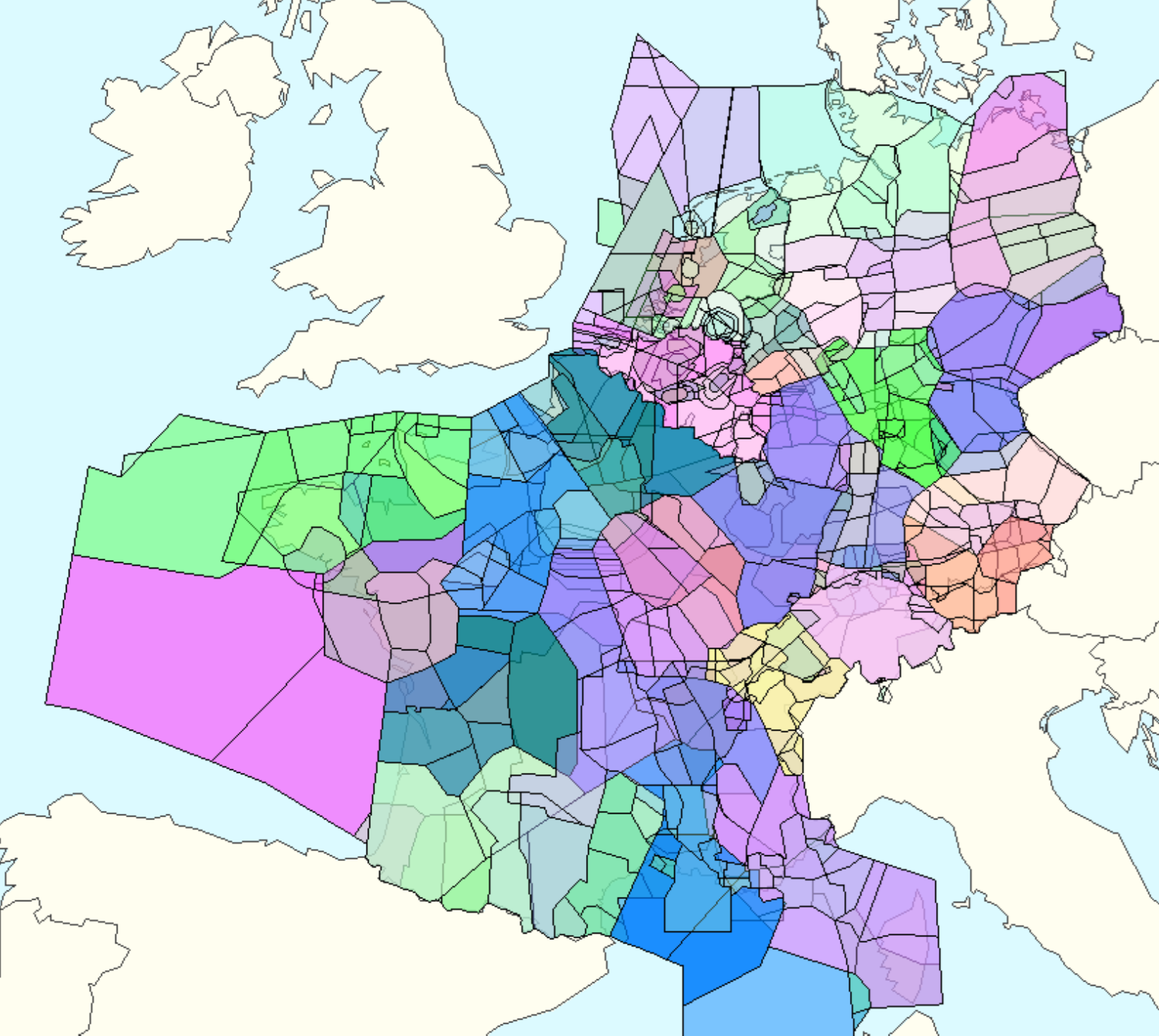}
    \caption{European Airspace - Sectors of FABEC}
    \label{fig:airspace}
\end{figure}

The heterogeneity affects all operational levels of an ANSP. Depending on the operational unit size, the en-route operations are allocated to multiple Area Control Centers to cover a specific area. The type, volume, and 3-D shape of the airspace controlled by an ACC can be very different. In FABEC (FAB Europe Central), there are ACCs dealing only with the Upper Airspace (e.g.,\ Maastricht Upper Airspace Control, MUAC), only with Lower Airspace (e.g., Munich ACC), and both airspaces (Lower and Upper, e.g., Zurich ACC), leading to a high degree of heterogeneity with regards to tools, systems and procedures \citet{FABEC2019}. ACCs are, in turn, divided into sector groups (licensed areas) and these into sectors. \autoref{fig:airspace} shows a map of the sector structure in Europe. DFS provided the figure based on the NEST tool \citep{EUROCONTROL2018h}.\

From an economic perspective, Air traffic control units decide on inputs, particularly air traffic control officers (ATCOs), and produce the relevant outputs (e.g., flight hours). Benchmarking is only helpful if the benchmarked unit has some decision power or if the principal decides on the allocation of inputs. The latter is currently not the case in Europe due to the country-based structure of air traffic control. Nevertheless, valid benchmarking must build upon an appropriate model that considers the markets' operational constraints. It has further to consider that ANSP-Services are divided into `Terminal’ and `En-route‘ operations. The operations differ significantly in terms of procedures and tasks. Since these services represent the main output of an ANSP, performance benchmarking must include both.\

\subsection{General Approach and Methods}

The performance benchmarking of companies is a central element in economics and business administration. There are numerous studies on methods~\citep{Fried2008, Lovell1992, Stepan2014} and applications~\citep{Ahmada2017, Albrecht2012, Hoffmann2006} in the various sectors of the economy. Performance benchmarking in aviation, especially in the ANSP context, on the other hand, is still a young discipline.\

In recent years, various methods have been established to compare companies' and enterprises' performance. Key figures or index figures represent a quotient of output (e.g., flight hours) and input (e.g., ATCO hours). In other words, the output is set in relation to the resources used. The higher this ratio, the higher the productivity or efficiency of the company. The advantage is the intuitively simple interpretation through the absolute productivity values and the rankings based on them. For instance, EUROCONTROL uses this methodology in its official reports.\

The main disadvantage if the above method is the limitation on one input and one output. Substituting one of the factors might lead to completely different results (performance scores and rankings) and thus to a lack of robustness. This issue can be solved by generating production or cost functions, enabling one to consider multiple inputs and outputs. Since a functional relationship between resources (or costs) and produced goods cannot be analytically derived in most cases \citep{Bielecki2011}, marginal production functions are usually empirically determined either by parametric or non-parametric approaches \citep{Fried2008}. \

Using parametric methods, ex-ante assumptions have to be made about the functional form of the production function. Non-parametric methods do not require ex-ante assumptions about functional relationships or disturbance variables that are instead determined by mathematical programming. In this case, efficiency influencing factors are integrated into the analysis via econometric methods, such as regression analyses, as a second step \citep{Banker2008, Hof2007, Simar2007}. Since an ex-ante assumption of functional relationships is challenging, and there is the risk of model miss-specification, we argue that the application of the deterministic, non-parametric Data Envelopment Analysis represents the most sufficient approach. The method is explained comprehensively, e.g., in \citet{A.Charnes1978}. 

Over the past decades, researchers have developed various DEA approaches. As one example, the super-efficiency DEA enables efficiency values of over 100\%. This approach has two main advantages: First, efficient units can also be ranked; second, this analysis helps to find outliers and oddities. Additive or multiplicative models, e.g., the slack-based DEA \citep{Tone2001}, combine input and output orientation \citep{Zhu2014}.  As a non-parametric approach, DEA provides no measures of model quality. Therefore, \cite{Bogetoft2011} developed the bootstrap DEA, a stochastically `corrected' or `adjusted' production function is generated. However, according to \citet{Coelli2005}, the bootstrap algorithm should not be applied in the case of empirically gathered data.

\section{The Economic Modeling}
\subsection{Processes and Value Creation Chain}
Like other companies, an ANSP uses various resources to provide its services. There are different monetary and non-monetary indicators available to measure ANSPs' inputs and outputs. However, since the service provision consists of many complex individual processes, it is first advisable to take a closer look at the value creation process.\

According to the definition, the actual productivity or efficiency value is the ratio of output and the production factors used for it. The resulting score reflects an ANSP's  performance either in absolute or relative terms. The output side comprises the core business area of air navigation service providers. Various operational (e.g., flight hours) and financial parameters (e.g., revenue) can represent the output. On the input side, personnel, technical equipment, software used, and bound capital are particularly relevant. This can be differentiated according to center (en-route) and tower (terminal) operations. 

In addition, endogenous (e.g., operational structure of the ANSP) or exogenous factors (e.g., geography) may influence productivity. A strict separation between -- or a clear division of -- the individual factors is impossible. For example, bound capital is interdependent with a country's legal foundations or with European law. Furthermore, interdependencies between traffic characteristics and the airspace division play an important role. Procedures (e.g., holdings) are primarily dependent on traffic demand. The complexity of traffic flows influences the airspace capacity and, thus, the number of feasible flights in terms of arrivals per hour and the "occupancy" value.\

Despite various attempts to standardize European systems and procedures, the ATM still features significant heterogeneity. Air navigation service providers differ in terms of services offered, legal form (e.g.,\ joint-stock company), or ownership (state-owned or partially privatized). The economic modeling should consider these differences since they are expected to influence performance. However, these characteristics are mostly exogenous and are, therefore, only to be taken into account in the second stage, but not in benchmarking.\

\subsection{Inputs and Outputs: Get rid of money}
Production factors mainly comprise human resources (HR), materials, capital, energy, and purchased services. The input can be expressed either by quantities or costs. However, monetary values are inappropriate due to the pan-European heterogeneity in price levels and exchange rates. Input costs vary significantly across Europe. In 2017, the annual employment cost per air traffic controller ranged from \EUR{}17,894 in Ukraine to \EUR{}277,629 at LVNL, ANSP in the Netherlands. That means that the annual cost per air traffic controller in the Netherlands is about 16 times higher than that of the Ukrainian ANSP UkSATSE. The cost per ATCO hour ranged from \EUR{}12 (UkSATSE) to \EUR{}232 (DFS / Germany). The difference in costs is due in particular to the differences in purchasing power between the European countries.\

These differences emphasize the need to avoid monetary values. Using costs as inputs would lead to the statement, that the most efficient ANSP is that with the lowest input prices. However, this, in turn, is largely dependent on the wages (e.g., for the controllers), which are primarily exogenous. Hence, when using costs, an ANSP would be evaluated with respect to indicators that it cannot influence, in other words, without considering operational and economic characteristics. This has been criticized concerning the EUROCONTROL reports and some academic papers. The authors sometimes use purchasing power parities to `adjust' costs between countries, however, it is not clear whether ATCO-hours' wage differences are systematically related to purchasing power differences. Therefore, we avoid using wages.\

The most crucial resource is the air traffic controllers. Due to different working time models, we use the number of controllers in Full-Time Equivalents (FTEs). This means that the number of controllers is aggregated to FTEs via a correction procedure. In addition to this pure number of (full-time) positions, working hours are recorded (ATCO-hours). Controllers may also work on projects or act as trainers. Therefore, both inputs are differentiated into further subcategories \citep{EUROCONTROL2008a,EUROCONTROL2012a,ICAO2018}. Besides air traffic controllers, air navigation service providers also employ personnel for administration, maintenance of technical equipment, and various other areas. Again, the differentiation between the number of FTEs and the person-hours expended applies.\

Bound capital includes buildings, technical equipment, and facilities, and other capital goods. They depend, to a large extent, on the organizational structure of the ANSP. If an ANSP operates several ACCs, this has an impact on the maintenance costs of the building, its energy requirements, and personnel expenses. The operation of towers also has an equivalent impact on capital.\

Analogous to inputs, outputs can be measured in monetary or operational terms. Operational outputs, in turn, are either quantity-based or time-based. Potential quantities are the number of IFR flights or flight hours serviced, IFR kilometers flown, IFR flight movements at airports and the number of composite flight hours or revenue.\ 

Air traffic control provides services for en-route flights and in the terminal area (takeoffs and landings). The corresponding outputs are represented, for example, by the controlled IFR flight hours $f$ and the IFR flight movements at the airports $a$. To calculate a performance indicator for a gate-to-gate consideration, a combined measure for both output parameters was introduced: The Composite Flight Hours (CFH). This value is the weighted sum of both output parameters \eqref{CFH1}. The weighting factor $w$ depends on terminal service costs ($TC$) and en-route service costs ($EC$). It is the pan-European ratio between the terminal and en-route unit costs $UC$, illustrated in formula \eqref{CFH2}. The PRU uses these values to rate air navigation services according to cost efficiency (\EUR{} per CFH) or productivity (CFH per ATCO hour) \citep{EUROCONTROL2019g}.\

\begin{equation}
  CFH=f+w \cdot a \label{CFH1}
 \end{equation}

\begin{equation}
  w=\frac{\frac{TC}{a}}{\frac{EC}{f}}=\frac{UC(a)}{UC(f)} \label{CFH2} = 0.27
 \end{equation}

Although there are some advantages to using a combined, uniform output measure, both the weighting value and the resulting output represent an artificial quantity. Thus, it represents a rough approximation of the total output. Due to the heterogeneity of ANSPs, a pan-European value for the weighting (0.27, see \cite{EUROCONTROL2020i}) may not be useful since individual weights show a strong dispersion. As argued in \citet{Standfuss2018e}, it might be useful to use individual cost shares to weight the CFH.\

After looking at potential factors, we present some selected data highlighting pan-European heterogeneity. EUROCONTROLs Performance Review Unit (PRU) provided the via the Onesky Portal, called `ACE data'. The database comprises about 60 operational and monetary indicators, some distinguished into en-route, terminal, and gate-to-gate services. The PRU databases currently include up to 38 ANSPs for 2003-2019. Table \ref{table:1} shows descriptive statistics for some selected data points. 

\begin{table}[htb!]
\centering \footnotesize
\begin{tabular}{lrrrrr}
 \hline \hline
Indicator & Min & Median & 3rd Quartile & Max & Std.Dev.\\
 \hline
Airspace Size (km$^2$) & 20.400 & 151.500 & 560.250 & 2.190.000 & 428.848 \\
ACCs &1 &1 &2 &5 &1 \\
Tower &0 &6 &16 &77 &14\\ 
IFR Flights &38.968 &611.342 &860.928 &3.015.153 &746.008\\ 
IFR Flight Hours & 9.442 & 237.314 & 426.481 & 2.287.512 & 511.051\\ 
Airport Movements & 0 &161.381 &479.862 &2.017.084 &545.210 \\
Composite Flight Hours &15.302 &286.178 &545.122 &2.777.883 &648.698\\ 
 \hline
\end{tabular}
\caption{Descriptive Statistics 2016}
\label{table:1}
\end{table}

As shown in the table, the airspace of Spain is about twice the size of the second-largest (France) and about 107 times larger than the smallest airspace (Slovenia). These differences have a partial impact on overall demand: France is the ANSP with the most controlled flight hours and flights: In terms of flights, demand is 77 times greater than in Armenia, and in terms of flight hours, it is even 242 times greater than the one of Moldova.\

\begin{result}
    Use quantitative inputs and outputs.
\end{result}

\subsection{Selection of ANSPs: Maastricht - in or out?}
One specific issue with benchmarking is to include comparable units only. Hence, a task is to determine outliers. Concerning this, we analyzed European data in two different ways. First we compared the data provided by PRU and summarized in Table \ref{table:1} with regard to extreme values. Our second contribution is to use DEA approaches to identify outliers.

As discussed earlier, a characteristic of the European ATM is the substantial heterogeneity. This can also be observed in the available data. Table \ref{table:1} shows some extreme values, such as 0 movements at the airport, respectively 100\% overflights (not shown in the table), emphasizing that MUAC represents a particular case. This is because MUAC only handles en-route traffic. Upper airspace units are usually less complex due to the lower vertical traffic component. Thus, in most cases, a higher throughput is possible, increasing productivity. Indeed, MUAC is always in first place in the evaluations, be it EUROCONTROL reports or academic studies. However, this leads to the question of whether Maastricht is so specific that it it not comparable with the other units at the ANSP level.\ 

We evaluated different models (selected inputs and outputs) by applying super-efficiency DEA \citep{Zhu2014}. The results confirm the specificity of MUAC. The super-efficiency values range from 180\% to 357\% for constant returns to scale, and 220\% to 409\% for variable returns to scale. No other efficient unit achieves such high score. This again highlights the special role of MUAC.\ 

The consequences are significant. In particular, when DEA is applied under constant returns to scale, this dominance leads to a significant shift in the frontier function and a devaluation of all inefficient units. Due to the calculation methodology for scale efficiency, the score would be biased as well. Therefore it is strongly recommended to remove MUAC from the considerations.\

\begin{result}
    Compare what is comparable. Do not use Maastricht ANSP in a European ANSP benchmarking.
\end{result}

\section{Benchmarking of European ANSPs}
\subsection{Data Selection and Modeling}

We mainly need input and output data for air navigation service providers to perform an efficiency analysis. On the European level, several data sources are available for this purpose, which differ primarily in granularity and the addressed operational level. The lower the operational level, the more observations can be used for analysis. The operational data needed to address sectors or licensed areas are available using the EUROCONTROL NEST tool \citep{EUROCONTROL2018h}. However, this data is not publicly available. Furthermore, only a few characteristics (factors) are recorded. It is also doubtful whether sectors or groups of sectors can used as DMUs in the DEA.\ 

We, therefore, focus on the ANSP level. The ACE database contains many characteristics \citep{EUROCONTROL2020e}. However, only a maximum of 38 observation units (depending on the year) are available. It means that the number of characteristics considered in parallel is also limited. The literature review already showed that the number of observation units is decisive. According to our preliminary studies, the DEA model should never consist of more than four factors (inputs and/or outputs). These four factors should therefore represent the described value chains as best as possible.\ 

Since DEA is a deterministic method and, thus, does not include stochastic errors, there are also high demands on data quality. This applies in particular, but not exclusively, to the standard DEA. Preliminary studies have shown that the collection of ACE data is not homogeneous across ANSP. This is due to EUROCONTROL's requirements and internal processes and models of the ANSPs for the data collection on the other hand. Consequently, data quality is lacking for some factors and some years. However, we experienced an increased in data quality after 2007.\ 

We prefer a high number of inputs (relative to output), as the ANSPs can determined inputs. The main trade-off for an ANSP on the output side is capacity versus costs. Generally, the output could be controlled by the ANSP by generating delay. In a DEA model, this could be reflected as negative or reciprocal output. However, tactical delay generation is not common and usually a consequence of uninfluencable factors. Consequently, we will not consider this (potential) trade-off nor the output, that is, thus, primarily exogenous. For the inputs, we use staff and bound capital, which are the primary cost drivers of ANS operations. However, since we argued that monetary values should be avoided, we include a composite infrastructure unit ($CIU$) that comprises towers $t$ and ACCs $a$ in the same manner as the composite flight hours \eqref{CIU}. The weighting factor is consistent with the one for CFH shown in formula \eqref{CFH2}. 

\begin{equation}
  CIU=a+w \cdot t \label{CIU}
 \end{equation}

As discussed above, this weighting is debatable since the unit cost shares may differ significantly. Subsequently, we introduce CFH and CIU with individual cost-share weightings to better reflect the heterogeneity. The corresponding factor is designated with the index $i$. We propose six models to approach efficiency benchmarking. The first two models (model 1) consider three inputs and two outputs. The other models combine the two outputs to composite flight hours, either with the PRU weighting (model 2) or the individual weighting (model 3). To emphasize the particularity of Maastricht UAC, the models are distinguished into two sub-models: Model A considers all available ANSPs, and model B excludes MUAC. Table \ref{table:2} summarizes the considered models.

\begin{table}[htb!]
    \centering
    \begin{tabular}{l|llll|ll}
    \hline\hline
    Model   & \multicolumn{1}{l|}{1A}                                & \multicolumn{1}{l|}{1B}                               & \multicolumn{1}{l|}{2A}    & 2B    & \multicolumn{1}{l|}{3A}                                    & 3B                                   \\ \hline
    Inputs  & \multicolumn{4}{l|}{\begin{tabular}[c]{@{}l@{}}ATCO hrs\\ Share Non-ATCOs \\ CIU\end{tabular}}                                                      & \multicolumn{2}{l}{\begin{tabular}[c]{@{}l@{}}ATCO hrs \\ Share Non-ATCOs \\ CIU\_i\end{tabular}} \\ \hline
    Outputs & \multicolumn{2}{l|}{\begin{tabular}[c]{@{}l@{}}Total Controlled Flight Hours\\ Airport Movements\end{tabular}} & \multicolumn{2}{l|}{CFH}           & \multicolumn{2}{l}{CFH\_i}                                                                        \\ \hline
    MUAC    & \multicolumn{1}{l|}{Incl.}                             & \multicolumn{1}{l|}{Excl.}                            & \multicolumn{1}{l|}{Incl.} & Excl. & \multicolumn{1}{l|}{Incl.}                                 & Excl.                                \\ \hline
    \end{tabular}
    \caption{DEA Models}
    \label{table:2}
\end{table}

Besides selecting factors for a suitable model, the applied methodology is crucial for the results. We focus on Data Envelopment Analysis since no apriori assumptions about functional relationships are required. Further, all DEAs are input-oriented. Nevertheless, the different types of DEA may lead to different results. Therefore, we will first discuss the methods analytically and then apply the various methods to verify the expectations empirically. \ 

\begin{result}
   Comprehensive data analysis is mandatory. Include data with appropriate accuracy in the DEA. When using ACE data, exclude years before 2008 from the database.
\end{result}

\subsection{Analytical Approach}
Based on the selection of factors it is useful to perform a standard DEA first. This should be carried out under the assumption of constant (CRS) as well as variable returns to scale (VRS). The results serve as a basis for comparison with the alternative methods.\ 

Applying the standard DEA for several years will most probably lead to volatility in the DEA scores, especially assuming CRS. It is necessary to check whether the volatility is due to fluctuations in the output and insufficient adjustment of resources by the DMU. In this case, the application of standard DEA is valid. Otherwise, there might be a data error.\ 

If some errors are not due to inputs or outputs it may be useful to bootstrap to evaluate the DEA model statistically. However, one should note that \citet{Coelli2005} particularly critizise the bootstrapping method. In the case of empirically observed data, the algorithm would lead to an artificial downgrade of efficiency scores. Since we use actual data which was recorded by the ANSPs, the results of the Bootstrap-DEA may not be robust.\

Super-efficiency DEA is primarily important to identify extreme values. We already created models excluding Maastricht, which is expected to be the most influencing DMU on the production function. Shifting the function will lead to lower efficiency scores for the ANSPs of the assigned peer-group. Further, the method might be helpful to re-evaluate all efficient units.\ 

Slack-Based DEA belongs to additive models, combining input minimization and output maximization. However, since ANSPs do not influence the output, the application might be seen as a model miss-specification. We expect biased and non-robust values in the DEA scores.\ 

We will check in the next section whether the assumptions can be confirmed. For this purpose, all discussed DEA types are applied for all years between 2008 and 2018 (due to data quality) and all available DMUs (ANSPs). We further distinguish the results into CRS-DEA and VRS-DEA.\

\subsection{Empirical Verification}

\begin{figure}[htb!]
    \centering
    \includegraphics[width=\textwidth]{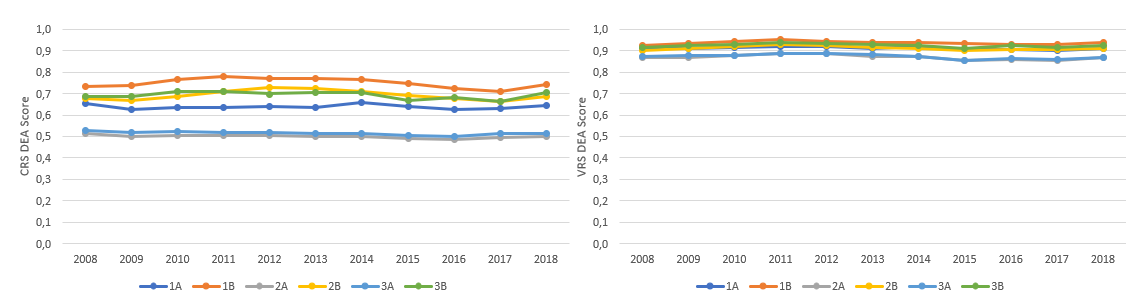}
    \caption{Comparison of average Standard-DEA scores between the six models}
    \label{fig:DeaModels}
\end{figure}

Based on all available data and models, we create and analyze more than 528 solution tables comparing the efficiency of ANSPs (6 models x 2 returns to scale types x 11 years x 4 DEA Types). Further, we also determine the ranks based on the calculated efficiency. It is impossible to show all results, but a selection of findings only. All other results are available on request. The comprehensive dataset enables us to compare the results with regard to different criteria. For example, \autoref{fig:DeaModels} shows the average DEA score (standard DEA) for all considered years. \ 

The figure shows that Model 1B provides  the highest efficiency score. The intuition for this outcome is that we use five factors implying that DEA classified more DMUs as efficient than in models using four factors. Further, since we exclude MUAC, the ANSPs in the MUAC peer group achieve higher efficiency scores. The latter is also visible for all other models: The B-versions always show significantly higher DEA scores than the A-variants. \

We can also observe that these differences decrease when implementing variable returns to scale. Overall, the scores are higher since the assumption of VRS leads to a convex production function. Hence, the distance of inefficient units to the efficient frontier is lower. Please note that despite arguing to use 2008-2011 data only, we show all years, proving that also (potentially) inaccurate data may not hamper the overall picture.\ 

\begin{figure}[htb!]
    \centering
    \includegraphics[width=.8\textwidth]{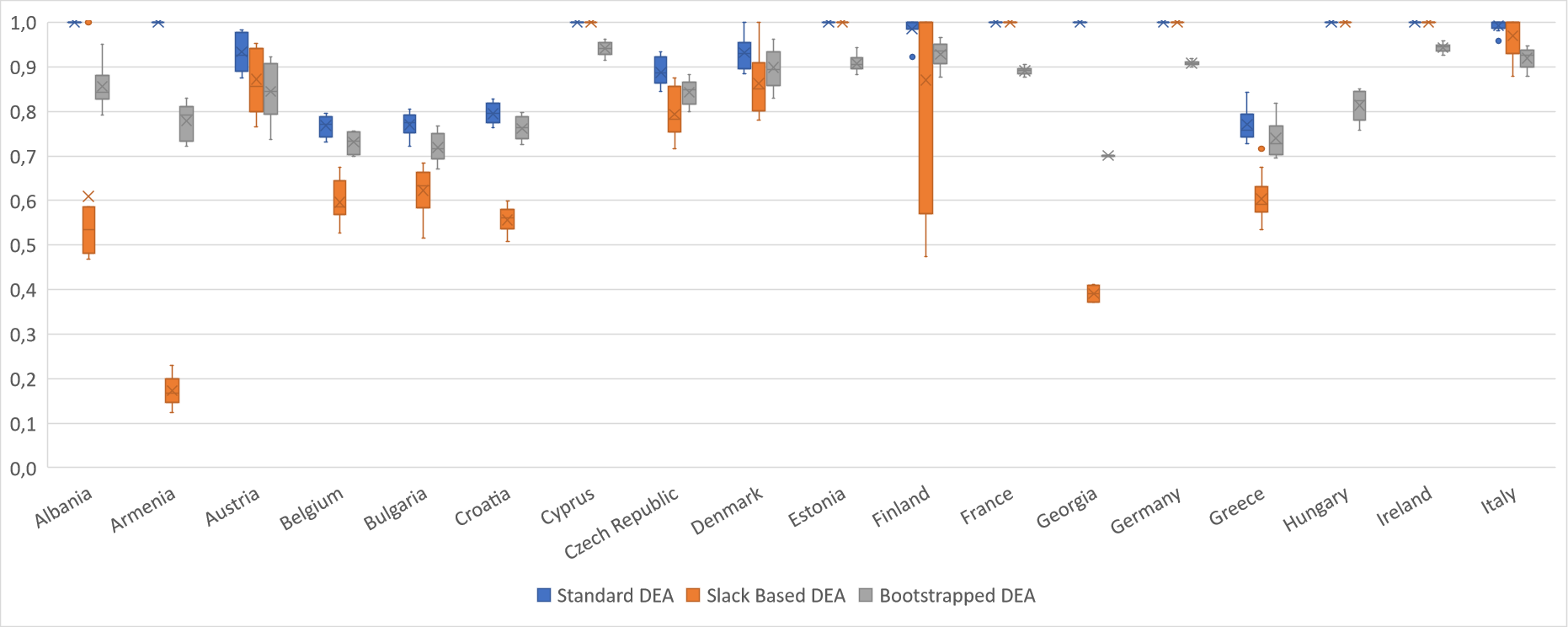}
    \caption{Spread of observed DEA efficiency scores according to the different types (1)}
    \label{fig:DTA}
\end{figure}

\begin{figure}[htb!]
    \centering
    \includegraphics[width=.8\textwidth]{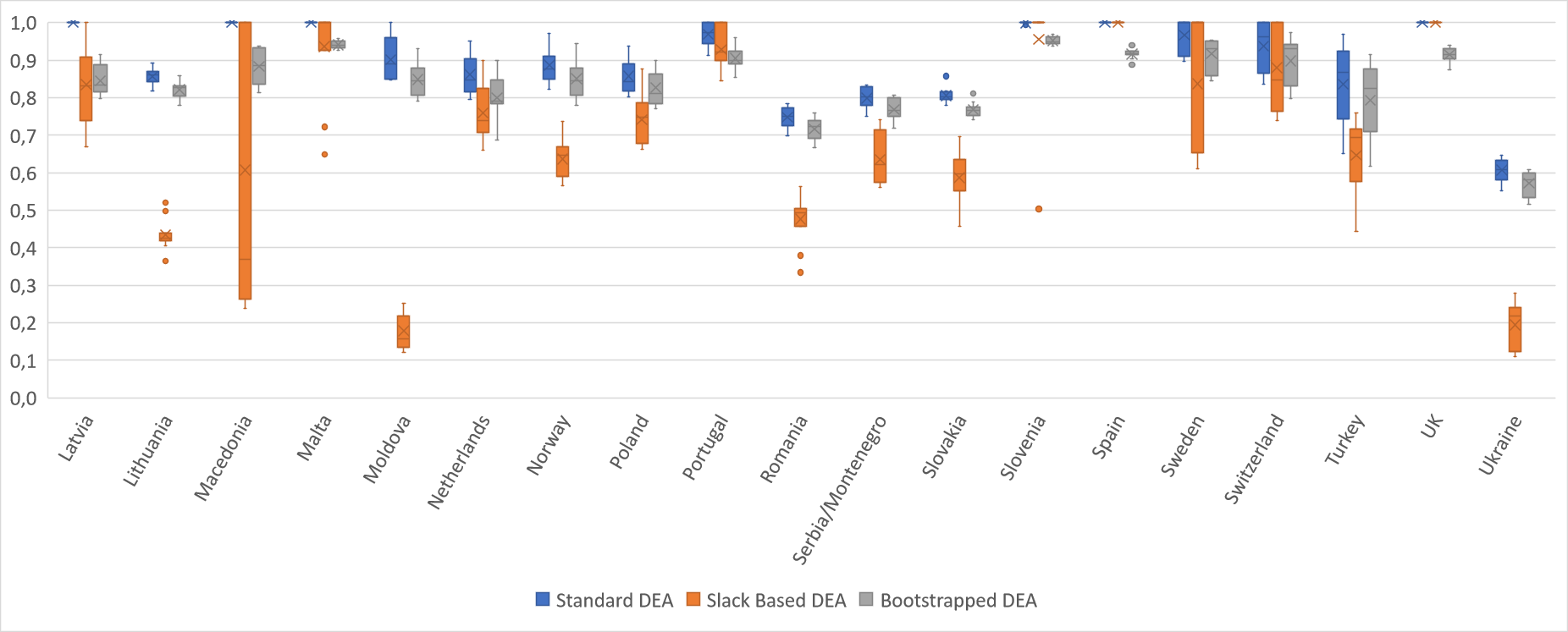}
    \caption{Spread of observed DEA efficiency scores according to the different types (2)}
    \label{fig:DTB}
\end{figure}

As discussed in the previous chapter, we need to check whether the scores of a method show a high spread. Therefore, we created boxplots showing the dispersion of the scores. The broader the box, respectively, the higher the span between the whiskers, the higher the volatility. \autoref{fig:DTA} and \autoref{fig:DTB} show the boxplots for three DEA-Types and all ANSPs, using 20018-2018 Data and implying variable returns to scale.\

A high range can also be caused by a trend. For example, an efficiency could increase annually by a few percentage points, so that after 11 years a high spread is shown in the boxplot. Therefore, we checked the annual distribution of the values in case of a large spread to analyze if the scores follow a trend or fluctuate significantly. The latter would prove the low robustness of the scores.

These figures help evaluate the different DEA types concerning robustness, respectively volatility in the scores. Applying a slack-based DEA is not helpful since scores are unstable. Thus, the results meet the expectations. Further, the scores of the bootstrapped DEA are almost like the scores of the standard DEA, reduced by some points. This is as expected since we used observed data. Thus we proved the statement by \citet{Coelli2005} and refute \citet{Button2013}.\ 

With regard to the super-efficiency DEA, we find the application quite reasonable. Depending on the model, several DMUs have an efficiency value exceeding unity. The very special role of MUAC is shown by efficiency values of up to over 350\%. In contrast, the exclusion of MUAC leads to values of 140\% maximum. France and Norway, in particular, achieve higher efficiency scores. \

\begin{figure}[htb!]
    \centering
    \includegraphics[width=.75\textwidth]{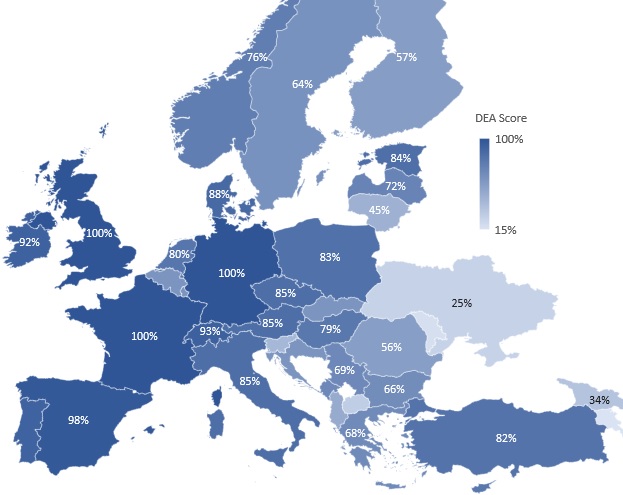}
    \caption{Average Efficiency Scores in Europe, 2008-2018}
    \label{fig:Map}
\end{figure}

We compare the DEA values at the European level based on our findings. For this purpose, we use model 2B (model 3B leads to similar results). \autoref{fig:Map} shows the average efficiency score of the ANSPs based on the standard CRS-DEA. The darker the shade, the more efficient the DMU. For illustration reasons, we assing the results to the respective countries and drop some designations (values). 

The figure shows that some ANSPs achieve efficiency in all considered years, leading to an average DEA score of 100\%. That mainly concerns large ANSPs in the European core area, such as DFS (Germany), DSNA (France), and NATS (UK). In contrast, ANSPs in the southeastern periphery are characterized by relatively low efficiency scores. Moldavia show the lowest efficiency, which the political conflicts in Ukraine and the overall small size of the airspace might cause. 

\begin{result}
    Standard DEA is feasible for evaluating the performance of ANSPs in Europe. Super-efficiency models lead to further insights, particularly regarding outliers.
\end{result}

\begin{result}
    Slack-Based DEA is not applicable in ANSP context due to missing robustness. Bootstrapping the DEA scores might provide an artificial error in the results.
\end{result}

\section{Conclusions}
In our study, we used different types of Data Envelopment Analysis to evaluate and compare the performance of European air navigation service providers. A meaningful benchmarking in the specific environment of ANSPs requires several important specifications.

First, an intensive data analysis shall identify outliers. If these extreme values are based on data errors or on the special nature of the DMU, the unit should be dropped. Here, the finding is that Maastricht Upper Airspace Control is a special case that should not be integrated into a comparative benchmarking analysis. After excluding this outlier, the case for choosing an appropriate DEA approach changes.\ 

Second, after comparing six models for performance benchmarking of ANSPs, we conclude that the DEA shall use a maximum of four factors (sum of inputs and outputs). Otherwise, the number of efficient DMUs is much higher, which hampers a root-cause analysis in the second stage. 

Third, input quantities shall be used instead of input costs in cross-country benchmarking since input costs are hardly comparable across countries. Hence, we  use operational determinants in the cross-country benchmarking. The results were processed concerning both constant and variable returns to scale.\ 

We have shown that the Standard DEA is appropriate for European ANSP benchmarking. Although the scores are partly subject to annual fluctuations, the results are robust and plausible in the majority of observations. In contrast, Slack-Based DEA and Bootstrap DEA are less suited for calculating the efficiency of ANSPs. This might be due to the low number of observations. As observed by the elimination of MUAC, the reduction of extreme heterogeneity lowers the case applying specific DEA approaches. We conclude that sophisticated DEA approaches hardly lead to any advantage compared to standard DEA concerning the accuracy of results or the addition of findings. However, super-efficiency can be help extend the results, especially when comparing efficient units.\

This paper focuses on annual data and compares (uni-periodic) DEA types. Since data is available for many years, a Malmquist analysis would be conceivable to investigate time effects such as the shift of production frontier. The method enables calculating and evaluating individual efficiency gains and losses for each unit. Alternatively, specifications of the DEA models could lead to further insights. That is left to future research. Further research may also study smaller operating units as discussed by \citet{Standfuss2017b}. That applies in particular to the ACCs, since they have specific decision-making power, e.g., concerning staffing).

The efficiency analysis is only the first step. The second stage would be a performance evaluation aiming to improve the efficiency of ANSPs. Therefore, future research will examine which endogenous and exogenous factors influence the DEA scores and which methods are suitable to quantify these influences. That will be studied in a second-stage analysis, as i.a. described in \cite{Hof2007}. The aim is to derive dedicated actions for ANSPs, regulators, or airspace users.

\section*{Acknowledgement}
The authors would like to thank Frank Fichert (UAS Worms) for his methodological support, as well as FABEC and DFS for their operational expertise, particularly Matthias Whittome, Thomas Hellbach, Christoph Czech and Juan Espinar Nova. We further thank the participants of the ITEA Conference 2022 in Toulouse, whose feedback contributed significantly to the improvement of the paper.

\bibliographystyle{apalike}
\bibliography{ref}

\end{document}